\documentclass{elsart}

\usepackage{amsmath,amssymb}
\allowdisplaybreaks

\numberwithin{equation}{section}

\journal{Annals of Physics}
\newcommand{\cA}{\mathcal{A}}
\newcommand{\cB}{\mathcal{B}}

\newcommand{\cH}{\mathcal{H}}

\newcommand{\cN}{\mathcal{N}}

\newcommand{\cU}{\mathcal{U}}
\newcommand{\cV}{\mathcal{V}}
\newcommand{\cW}{\mathcal{W}}
\newcommand{\bA}{\boldsymbol{A}}
\newcommand{\bB}{\boldsymbol{B}}
\newcommand{\bC}{\boldsymbol{C}}

\newcommand{\bS}{\mathbf{S}}
\newcommand{\tH}{\tilde{\!H}{}}

\newcommand{\tcH}{\tilde{\cH}}
\newcommand{\tcV}{\tilde{\cV}}

\newcommand{\bi}{\bar{\imath}}
\newcommand{\bj}{\bar{\jmath}}
\newcommand{\bk}{\bar{k}}
\newcommand{\bl}{\bar{l}}
\newcommand{\bmu}{\bar{\mu}}
\newcommand{\bbZ}{\mathbb{Z}}
\newcommand{\bbR}{\mathbb{R}}
\newcommand{\bbC}{\mathbb{C}}
\newcommand{\ii}{\mathrm{i}}
\newcommand{\ee}{\mathrm{e}}
\newcommand{\dd}{\mathrm{d}}
\newcommand{\del}{\partial}

\newcommand{\rnu}{\sqrt{\nu}}
\newcommand{\Lsl}{\mathfrak{sl}}

\newcommand{\sn}{\operatorname{sn}}
\newcommand{\cn}{\operatorname{cn}}
\newcommand{\dn}{\operatorname{dn}}
\begin{document}

\begin{frontmatter}

\title{A Many-body Generalization of Quasi-solvable Models
 with Type C $\cN$-fold Supersymmetry (I) Regular Cases}
\author{Toshiaki Tanaka} \ead{ttanaka@ciruelo.fis.ucm.es}
\address{Departamento de F\'{\i}sica Te\'orica II,
  Facultad de Ciencias F\'{\i}sicas,\\
  Universidad Complutense, 28040 Madrid, Spain}



\begin{abstract}

We make a generalization of the type C monomial space of a single variable,
which was introduced in the construction of type C $\cN$-fold supersymmetry,
to several variables. Then, we construct the most general quasi-solvable
second-order operators preserving this multivariate type C space.
These operators of several variables are characterized by the fact that two
different polynomial type solutions are available. In particular,
we investigate and classify all the possible Schr\"odinger operators realized
as a subclass of this family. It turns out that the rational, hyperbolic,
and trigonometric Calogero-Sutherland models as well as some particular type
of the elliptic Inozemtsev system, all associated
with the $BC_M$ root system, fall within the class.

\end{abstract}

\begin{keyword}
  quantum many-body problem\sep quasi({}-exact) solvability\sep
  Calogero-Sutherland models \sep Inozemtsev models\sep supersymmetry
 \PACS 03.65.Ge\sep 02.30.Jr
\end{keyword}


\end{frontmatter}

\section{Introduction}
\label{sec:intro}

Finding exact solutions for an equation describing a physical system is one
of the most important issues in mathematical physics. For quantum mechanical
systems, the discovery of the underlying Lie-algebraic structure in
Ref.~\cite{Tu88} has stimulated systematic construction of
(quasi-{})solvable models.
The first stage of the generalization of the idea to quantum many-body
systems was however confronted with the difficulty due to the fact that
second-order differential operators are in general not equivalent to
Schr\"odinger operators, see, e.g., Refs.~\cite{Sh89,GKO94,Tu94,Us94}.
The next stage began with the success \cite{RT95} of an algebraization of
the known exactly solvable Calogero-Sutherland (CS) models~\cite{Ca71,Su71}.
The key ingredient in Ref.~\cite{RT95} is to construct the generators of
$\Lsl(M+1)$  Lie algebra in terms of the elementary symmetric polynomials
which reflect the permutation symmetry of the CS models. The similar methods
were employed to show the (quasi-{})solvability of the variety of the CS
models~\cite{MRT96,BTW98,HS99,GGR00}. Finally, the complete classification
of the models constructed from the $\Lsl(M+1)$ algebraization was given
in Ref.~\cite{Ta04}. It turns out that all the obtained models are of
the Inozemtsev type, which was shown to be classically
integrable~\cite{In83,IM85,In89}. One of the key observations in
Ref.~\cite{Ta04} is the significance of the underlying symmetry of the
solvable sector, namely, the subspace preserved by the Hamiltonian.
Another aspect of the significant role of the symmetry in the one-body case
was reported in Ref.~\cite{Ta03a} in connection with $\cN$-fold supersymmetry.

In a previous paper~\cite{GT04a}, we developed a systematic algorithm to
construct an $\cN$-fold supersymmetric system starting from a function space
preserved by one of the pair Hamiltonians. We then applied it to monomial
type spaces to find out a new family of $\cN$-fold supersymmetry called
type C. As a byproduct, we found that for all finite values of the spectral
parameter the characteristic equation of the Hamiltonians with type C
$\cN$-fold supersymmetry possesses two linearly independent series
solutions around the origin in a suitable variable (cf. Section 5 in
Ref.~\cite{GT04a}). In this respect, there is a famous mathematical theorem,
called Fuchs's theorem, which tells a sufficient condition for second-order
linear differential operators of a single variable to acquire two linearly
independent series solutions. The applicability of the theorem however is
relatively limited and a generalization to partial differential equations
is in general difficult. Therefore, the procedure
presented in Ref.~\cite{GT04a} has a potential possibility to be quite
efficient in investigating what kind of linear differential operators,
especially in several variables, can have two or more linearly independent
series or polynomial solutions. In this context, it is natural to inquire
in particular whether a similar approach could be applied to investigate
what types of quantum many-body Hamiltonians admit the above property.

In this article, we will show that we can indeed systematically construct
a family of quantum many-body systems with this characteristic by a suitable
generalization of the one-body case in Ref.~\cite{GT04a}.
A key ingredient is how to generalize the so called
\emph{type C monomial space} of a single variable, which is
preserved by the type C $\cN$-fold supersymmetric Hamiltonians, to
several variables. This is actually non-trivial problem because there are
so many possible generalizations and we can hardly know
\emph{a priori}
which one is preferable to others. For this purpose, we employ significant
properties of the type C space of a single variable discussed in
Ref.~\cite{GT04a} as criteria.
It turns out that the criteria really work advantageously
in several contexts. In particular, we can mostly avoid the problem caused
from the fact that second-order differential operators are in general not
equivalent to Schr\"odinger operators, and we can completely keep the same
classification scheme as that in the single-variable case.

The article is organized as follows. In the next section, we summarize
some definitions related to quasi-solvability in order to avoid ambiguity.
In Section~\ref{sec:gentC}, we show how to generalize
the type C monomial space of a single variable to several variables.
Then in Section~\ref{sec:const} we construct the most general
second-order linear differential operator which preserves the generalized
type C space. The obtained operators are however not equivalent to a
Schr\"odinger operator in general. Thus we present in Section~\ref{sec:reduc}
a systematic procedure to extract the operators which are equivalent to
a Schr\"odinger operator, namely, gauged Hamiltonians from the previously
obtained family of the operators.
In Section~\ref{sec:inver}, we clarify how the gauged Hamiltonians obtained
in Section~\ref{sec:reduc} are embedded in the general type C operators
by identifying the relation between the free parameters in the operators.
Utilizing the form-invariance under the projective transformations,
we fully classify the possible Hamiltonians preserving
the generalized type C space in Section~\ref{sec:class}.
The explicit form of the potential as well as the solvable sectors are
exhibited in each case. In Section~\ref{sec:diffi}, we briefly discuss
possibility and difficulty in the generalization of the type B monomial
space to several variables. Finally, we summarize the results obtained
here and discuss further possible developments of the present work in
Section~\ref{sec:discus}.
Some useful formulas are summarized in Appendix.

\section{Definition}
\label{sec:defin}

First of all, we shall give the definition of quasi-solvability and
some notions of its special cases based on Refs.~\cite{Tu94,Ta04}.
A linear differential operator $H$ of several variables
$q=(q_{1},\dots,q_{M})$ is said to be \textit{quasi-solvable}
if it preserves a finite dimensional functional space $\cV_{\cN}$
whose basis admits an explicit analytic form $\phi_i(q)$:
\begin{align}
\label{eq:defqs}
H\cV_{\cN}&\subset\cV_{\cN}, & \dim\cV_{\cN}&=n(\cN)<\infty,
 & \cV_{\cN}&=\bigl\langle\phi_{1}(q),\dots,\phi_{n(\cN)}(q)
 \bigr\rangle.
\end{align}
An immediate consequence of the above definition of quasi-solvability
is that, since we can calculate finite dimensional matrix elements
$\bS_{k,l}$ defined by,
\begin{align}
\label{eq:defbS}
H\phi_{k}=\sum_{l=1}^{n(\cN)}\bS_{k,l}\phi_{l}\qquad
 \bigl(k=1,\dots,n(\cN)\bigr),
\end{align}
we can diagonalize the operator $H$ and obtain its spectra
in the space $\cV_{\cN}$, at least, algebraically. Furthermore,
if the space $\cV_{\cN}$ is a subspace of a Hilbert space
$L^{2}(S)$ $(S\subset\bbR^{M})$ on which the operator $H$ is naturally
defined, the calculable spectra and the corresponding vectors in
$\cV_{\cN}$ give the \textit{exact} eigenvalues and eigenfunctions of
$H$, respectively. In this case, the operator $H$ is said to be
\textit{quasi-exactly solvable} (on $S$). Otherwise, the calculable
spectra and the corresponding vectors in $\cV_{\cN}$ only give
\textit{local} solutions of the characteristic equation of $H$.

A quasi-solvable operator $H$ of several variables
is said to be \textit{solvable}
if it preserves an infinite flag of finite dimensional functional
spaces $\cV_{\cN}$,
\begin{align}
\label{eq:flagv}
\cV_{1}\subset\cV_{2}\subset\dots\subset\cV_{\cN}\subset\cdots,
\end{align}
whose bases admit explicit analytic forms, that is,
\begin{align}
\label{eq:defsv}
H\cV_{\cN}&\subset\cV_{\cN}, & \dim\cV_{\cN}&=n(\cN)<\infty,
 & \cV_{\cN}&=\bigl\langle\phi_{1}(q),\dots,\phi_{n(\cN)}(q)
 \bigr\rangle,
\end{align}
for $\cN=1,2,3,\ldots$.
Furthermore, if the sequence of the spaces (\ref{eq:flagv}) defined
on $S\subset\bbR^{M}$ satisfies,
\begin{align}
\label{eq:limvs}
\overline{\cV_{\cN}(S)}\rightarrow L^{2}(S)\qquad
 (\cN\rightarrow\infty),
\end{align}
the operator $H$ is said to be \textit{exactly solvable} (on $S$).

\section{A Generalization of the Type C Monomial Space}
\label{sec:gentC}

In this article, we shall consider a quantum system of $M$ identical
particles on a line. The Hamiltonian is thus given by
\begin{align}
\label{eq:mbHam}
H=-\frac{1}{2}\sum_{i=1}^{M}\frac{\del^2}{\del q_i^2}+V(q_1,\dots,q_M)\,,
\end{align}
where the potential has permutation symmetry:
\begin{align}
V(\dots,q_i,\dots,q_j,\dots)=V(\dots,q_j,\dots,q_i,\dots)
 \qquad\forall i\neq j\,.
\end{align}
To construct a quasi-solvable operator of the form \eqref{eq:mbHam},
we shall follow the three steps after Ref.~\cite{Ta04}, namely, i) a
\emph{gauge} transformation on the Hamiltonian \eqref{eq:mbHam}:
\begin{align}
\label{eq:gtHam}
\tH=\ee^{\cW(q)}H\ee^{-\cW(q)}\,,
\end{align}
ii) a change of variables from $q_i$ to $z_i$ by a function $z$ of a single
variable:
\begin{align}
z_i=z(q_i)\,,
\end{align}
and iii) the introduction of the elementary symmetric polynomials of $z_i$
defined by,
\begin{align}
\label{eq:espol}
\sigma_k(z)=\sum_{i_1<\dots<i_k}^{M}z_{i_1}\dots z_{i_k}
 \qquad (k=1,\dots,M),\qquad\sigma_0\equiv 1\,.
\end{align}
Due to the permutation symmetry of the original Hamiltonian \eqref{eq:mbHam},
the gauged Hamiltonian \eqref{eq:gtHam} can be completely expressed in terms
of the elementary symmetric polynomials \eqref{eq:espol}.

The next task is to choose a vector space to be preserved by the gauged
Hamiltonian \eqref{eq:gtHam}. The type C monomial space of a single variable
$z$ is defined by~\cite{GT04a},
\begin{align}
\label{eq:stypC}
\tcV_{\cN_1\!,\cN_2}^{(C)}=\left\langle 1,z,\dots,z^{\cN_1-1},
 z^\lambda,z^{\lambda+1},\dots,z^{\lambda+\cN_2-1}\right\rangle\,,
 \qquad \cN=\cN_1+\cN_2\,,
\end{align}
where $\cN_1$ and $\cN_2$ are positive integers satisfying $\cN_1\geq\cN_2$
and $\lambda$ is a real number with the restriction
\begin{align}
\label{eq:resla}
\lambda\in\bbR\setminus\{-\cN_2,-\cN_2+1,\dots,\cN_1\}\,,
\end{align}
and with $\lambda\neq -\cN_2-1, \cN_1+1$ if $\cN_1=1$ or $\cN_2=1$.
It is however evident that there are many possibilities in generalizing
the space \eqref{eq:stypC} to several variables. Hence, we need
suitable principles in order to render the possible generalizations unique.
For this purpose, we shall recall the characteristic features of the type C
space \eqref{eq:stypC} discussed in Ref.~\cite{GT04a}. First of all,
the space \eqref{eq:stypC} decomposes as the direct sum of two spaces:
\begin{align}
\label{eq:sdecp}
\tcV_{\cN_1\!,\cN_2}^{(C)}=\tcV_{\cN_1}^{(A)}\oplus z^\lambda
 \,\tcV_{\cN_2}^{(A)}\,,
\end{align}
where $\tcV_{\cN_k}^{(A)}$ is the type A monomial space of dimension $\cN_k$
defined by
\begin{align}
\label{eq:stypA}
\tcV_{\cN_k}^{(A)}=\left\langle 1,z,\dots,z^{\cN_k-1}\right\rangle\,.
\end{align}
Thus, we shall generalize the type C space \eqref{eq:stypC} so that a
similar decomposition is realized. The next task is then to find a
several-variable counterpart of the type A space \eqref{eq:stypA}. In
this respect, we note the fact that the type A space \eqref{eq:stypA}
provides an $\Lsl(2)$ module, which is the foundations of the $\Lsl(2)$
construction of quasi-solvable models in Ref.~\cite{Tu88}.
Therefore, a natural generalization of the type A space to several
variables is provided by an $\Lsl(M+1)$ module given by
\begin{align}
\label{eq:mtypA}
\tcV_{\cN_k;M}^{(A)}=\Bigl\langle \sigma_1^{n_1}\dots\sigma_M^{n_M}:
 n_i\in\bbZ_{\geq 0}, 0\leq\sum_{i=1}^{M}n_i\leq\cN_k-1\Bigr\rangle\,.
\end{align}
Indeed, this space is exactly what the quasi-solvable $M$-body Hamiltonians
constructed in Ref.~\cite{Ta04} from $\Lsl(M+1)$ generators preserve, and
reduces to the type A space of a single variable \eqref{eq:stypA} when $M=1$. 
{}From the decomposition criterion, the generalized type C space we shall take
thus have the following form:
\begin{align}
\label{eq:mdecp}
\tcV_{\cN_1\!,\cN_2;M}^{(C)}=\tcV_{\cN_1;M}^{(A)}\oplus
 f_\lambda(z)\,\tcV_{\cN_2;M}^{(A)}\,.
\end{align}
The function $f_\lambda$ in Eq.~\eqref{eq:mdecp} must be $f_\lambda(z)
=z^\lambda$ when $M=1$ in order that the space \eqref{eq:mdecp} reduces to
Eq.~\eqref{eq:sdecp}. Next, we recall the fact that the type C space
\eqref{eq:stypC} is form-invariant under a special projective transformation.
More explicitly, the space \eqref{eq:stypC} satisfies
\begin{align}
\label{eq:sinvC}
\tcV_{\cN_1\!,\cN_2}^{(C)}[z,\lambda]\mapsto
 z^{\cN_1-1}\tcV_{\cN_1\!,\cN_2}^{(C)}[z^{-1},\lambda]
 =\tcV_{\cN_1\!,\cN_2}^{(C)}[z,\cN_1-\cN_2-\lambda]\,.
\end{align}
In Ref.~\cite{Ta04} it was shown that the invariance under projective
transformations of the vector space preserved by the Hamiltonians plays
crucial roles in constructing quasi-solvable quantum many-body systems.
Thus, we shall require that the generalized type C space has a similar
property to Eq.~\eqref{eq:sinvC}. To see how the space of the form
\eqref{eq:mdecp} behaves under the special projective transformation,
we first note that the elementary symmetric polynomials \eqref{eq:espol}
are transformed according to
\begin{align}
\label{eq:tfesp}
\sigma_k(z)\mapsto\sigma_k(z^{-1})=\sigma_{M-k}(z)\,\sigma_M(z)^{-1}\,.
\end{align}
{}From this formula, we easily see that the type A space of several variables
defined by Eq.~\eqref{eq:mtypA} has the following invariance property:
\begin{align}
\tcV_{\cN_k;M}^{(A)}[z]\mapsto\tcV_{\cN_k;M}^{(A)}[z^{-1}]
 =\sigma_M^{-(\cN_k-1)}\,\tcV_{\cN_k;M}^{(A)}[z]\,.
\end{align}
Hence, the space \eqref{eq:mdecp} transforms as follows:
\begin{align}
\tcV_{\cN_1\!,\cN_2;M}^{(C)}[z,\lambda]\mapsto&\,\tcV_{\cN_1\!,\cN_2;M}^{(C)}
 [z^{-1},\lambda]\notag\\
=&\,\sigma_M^{-(\cN_1-1)}\left(\tcV_{\cN_1;M}^{(A)}[z]\oplus
 \sigma_M^{\cN_1-\cN_2}f_\lambda(z^{-1})\,\tcV_{\cN_2;M}^{(A)}[z]\right)\,.
\end{align}
It is now clear that the space \eqref{eq:mdecp} has a suitable form-invariance
similar to Eq.~\eqref{eq:sinvC} if and only if the function $f_\lambda$ in
Eq.~\eqref{eq:mdecp} satisfies
\begin{align}
\label{eq:eqfla}
\sigma_M^{\cN_1-\cN_2}f_\lambda(z^{-1})=f_{\cN_1-\cN_2-\lambda}(z)\,.
\end{align}
Actually if this is the case, we have
\begin{align}
\label{eq:minvC}
\tcV_{\cN_1\!,\cN_2;M}^{(C)}[z,\lambda]\mapsto
 \sigma_M^{\cN_1-1}\tcV_{\cN_1\!,\cN_2;M}^{(C)}[z^{-1},\lambda]=
 \tcV_{\cN_1\!,\cN_2;M}^{(C)}[z,\cN_1-\cN_2-\lambda]\,,
\end{align}
which reduces exactly to Eq.~\eqref{eq:sinvC} when $M=1$. The solution of
Eq.~\eqref{eq:eqfla} is easily found if we note that the dependence on
$\cN_1-\cN_2$ only comes from the variable $\sigma_M$ and take the formula
\eqref{eq:tfesp} into account. Only the possible solution for $f_\lambda$,
which reduces to $z^\lambda$ when $M=1$, is given by
\begin{align}
f_\lambda(z)=\sigma_M(z)^{\lambda}\,.
\end{align}
In summary, we have generalized the type C monomial space \eqref{eq:stypC}
of a single variable to several variables by the requirement of
the decomposition structure \eqref{eq:mdecp} and of the form-invariance
under the special projective transformation \eqref{eq:minvC}, and finally
obtain
\begin{align}
\label{eq:mtypC}
\tcV_{\cN_1\!,\cN_2;M}^{(C)}=\tcV_{\cN_1;M}^{(A)}\oplus\sigma_M^\lambda
 \,\tcV_{\cN_2;M}^{(A)}\,.
\end{align}

\section{Construction of Quasi-solvable Operators}
\label{sec:const}

In this section, we shall construct the general quasi-solvable operators of
(at most) second-order leaving the type C space \eqref{eq:mtypC} invariant.
For the first-order differential operators, we have the following set of
independent ones:
\begin{subequations}
\label{eqs:1ops}
\begin{align}
\frac{\del}{\del\sigma_{\bi}}&\equiv E_{0\bi}\qquad (\bi=1,\dots,M-1)\,,\\
\sigma_i \frac{\del}{\del\sigma_{\bj}}&\equiv E_{i\bj}
 \qquad (i=1,\dots,M; \bj=1,\dots,M-1)\,,\\
\label{eq:1op02}
\sigma_M \frac{\del}{\del\sigma_M}&\equiv E_{M\!M}\,.
\end{align}
\end{subequations}
In the above and hereafter, we employ the convention that the Latin indices
take integer values from 1 to $M$ while the ones with bar take integer
values from 1 to $M-1$. We note that $E_{0M}$ and $E_{\bi M}$ cannot
preserve the elements $\sigma_1^{n_1}\dots\sigma_{M-1}^{n_{M-1}}
\sigma_M^\lambda$ inside the space $\tcV_{\cN_1\!,\cN_2;M}^{(C)}$ due to the
restriction on $\lambda$, Eq.~\eqref{eq:resla}. We also note that any
first-order operator which raises the degree cannot preserve the space
$\tcV_{\cN_1\!,\cN_2;M}^{(C)}$.

For the independent second-order differential operators, we have the
following set:
\begin{align}
\frac{\del^2}{\del\sigma_{\bk}\del\sigma_{\bl}}=E_{0\bk}E_{0\bl}
 \qquad (\bk\geq\bl)\,,
\end{align}
\begin{subequations}
\begin{align}
\sigma_i \frac{\del^2}{\del\sigma_{\bk}\del\sigma_{\bl}}&=E_{i\bk}E_{0\bl}
 \qquad (\bk\geq\bl)\,,\\
\sigma_M \frac{\del^2}{\del\sigma_M \del\sigma_{\bk}}&=E_{M\!M}E_{0\bk}\,,\\
\label{eq:2op0-3}
\frac{\del}{\del\sigma_M}\Bigl(\sigma_M \frac{\del}{\del\sigma_M}
 -\lambda\Bigr)&\equiv E_{M\!M,0M}\,,
\end{align}
\end{subequations}
\begin{subequations}
\begin{align}
\sigma_i \sigma_j \frac{\del^2}{\del\sigma_{\bk}\del\sigma_{\bl}}
 &=E_{i\bk}E_{j\bl}-\delta_{j\bk}E_{i\bl}\qquad (i\geq j\,;
 \bk\geq\bl)\,,\\
\sigma_M\sigma_i \frac{\del^2}{\del\sigma_M \del\sigma_{\bk}}
 &=E_{M\!M}E_{i\bk}-\delta_{iM}E_{M\!\bk}\,,\\
\label{eq:2op003}
\sigma_M^2\frac{\del^2}{\del\sigma_M^2}
 &=E_{M\!M}E_{M\!M}-E_{M\!M}\,,\\
\label{eq:2op004}
\sigma_{\bi}\frac{\del}{\del\sigma_M}\Bigl(\sigma_M\frac{\del}{\del\sigma_M}
 -\lambda\Bigr)&\equiv E_{M\!M,\bi M}\,
\end{align}
\end{subequations}
\begin{align}
\label{eq:2op+0}
\sigma_i \biggl(\cN_1-1-\sum_{k=1}^{M}\sigma_k \frac{\del}{\del\sigma_k}
 \biggr)\biggl(\lambda+\cN_2-1-\sum_{l=1}^{M}\sigma_l
 \frac{\del}{\del\sigma_l}\biggr)\equiv E_{i0,00}\,.
\end{align}
The operators \eqref{eq:2op0-3}, \eqref{eq:2op004}, and \eqref{eq:2op+0}
cannot be expressed as polynomials of the set of the first-order operators
\eqref{eqs:1ops}. In the case of $M=1$, only the four independent operators,
namely, Eqs.~\eqref{eq:1op02}, \eqref{eq:2op0-3}, \eqref{eq:2op003}, and
\eqref{eq:2op+0} with $i=M=1$ exist, which reproduces the result for the
type C quasi-solvable operators of a single variable given in
Ref.~\cite{GT04a}. We also note the important fact that all the operators
listed above preserve \emph{separately} both the subspaces of
$\tcV_{\cN_1\!,\cN_2;M}^{(C)}$ in Eq.~\eqref{eq:mtypC},
as in the case of $M=1$~\cite{GT04a}. The most general linear differential
operator of (at most) second-order which leaves the type C
space \eqref{eq:mtypC} invariant is thus given by
\begin{align}
\label{eq:gqsop1}
\tcH=&\,-\sum_i A_{i0,00}E_{i0,00}
 -\sum_{\mu\geq\nu,\bk\geq\bl}
 A_{\mu\bk,\nu\bl}E_{\mu\bk}E_{\nu\bl}-\sum_{\bmu}
 A_{M\!M,\bmu M}E_{M\!M,\bmu M}\notag\\
&\,-\sum_{\mu,\bk}A_{M\!M,\mu\bk}E_{M\!M}E_{\mu\bk}
 -A_{M\!M,M\!M}E_{M\!M}E_{M\!M}\notag\\
&\,+\sum_{\mu,\bk}B_{\mu\bk}E_{\mu\bk}+B_{M\!M}E_{M\!M}-c_0\,,
\end{align}
where the coefficients $A$, $B$ with indices and $c_0$ are real constants.
The Greek indices without(with) bar take integer values from 0 to $M$($M-1$),
respectively. The summation above is understood to take all the possible
integer values indicated by the corresponding type of indices. In terms of
the variables $\sigma_k$, the operator $\tcH$ is expressed as
\begin{multline}
\label{eq:gqsop2}
\tcH=-\sum_{\bk,\bl=1}^{M-1}\bigl[\bA_0(\sigma)\sigma_{\bk}\sigma_{\bl}
 +\bA_{\bk\bl}(\sigma)\bigr]\frac{\del^2}{\del\sigma_{\bk}\del\sigma_{\bl}}
 -\sum_{\bk=1}^{M-1}\bigl[2\bA_0 (\sigma)\sigma_{\bk}\\
+\bA_{M\!\bk}(\sigma)\bigr]
 \sigma_M \frac{\del^2}{\del\sigma_M \del\sigma_{\bk}}
 -\bigl[\bA_0 (\sigma)\sigma_M +\bA_{M\!M}(\sigma)\bigr]\sigma_M
 \frac{\del^2}{\del\sigma_M^2}\\
+\sum_{k=1}^M \bigl[(\cN+\lambda-3)\bA_0(\sigma)
 \sigma_k -\bB_k (\sigma)\bigr]\frac{\del}{\del\sigma_k}
 -\bC(\sigma)\,,
\end{multline}
where $\bA_0$, $\bA_{kl}$, $\bB_{k}$, and $\bC$ are polynomials of
several variables of at most second-degree given by
\begin{subequations}
\begin{align}
\label{eq:polA0}
\bA_0(\sigma)=&\,\sum_{i=1}^M A_{i0,00}\,\sigma_i\,,\\
\label{eq:polAkl}
\bA_{\bk\bl}(\sigma)=&\,\sum_{i\geq j}^M A_{i\bk,j\bl}\,\sigma_i\sigma_j
 +\sum_{i=1}^M A_{i\bk,0\bl}\,\sigma_i +A_{0\bk,0\bl}\qquad (\bk\geq\bl)\,,\\
\label{eq:polAMk}
\bA_{M\!k}(\sigma)=&\,\sum_{i=1}^M A_{M\!M,ik}\,\sigma_i+A_{M\!M,0k}\,,\\
\label{eq:polBk}
\bB_{\bk}(\sigma)=&\,\sum_{i=1}^M \biggl(\,\sum_{(i\geq)\bj(\geq\bk)}^{M-1}
 A_{i\bj,\bj\bk}-B_{i\bk}\biggr)\sigma_i +A_{M\!M,M\!\bk}\,\sigma_M
 -B_{0\bk}\,,\\
\label{eq:polBM}
\bB_M(\sigma)=&\,-(\lambda-1)\bA_{M\!M}(\sigma)
 +(\lambda A_{M\!M,M\!M}-B_{M\!M})\,\sigma_M\,,\\[10pt]
\label{eq:polbC}
\bC(\sigma)=&\,(\cN_1-1)(\cN_2+\lambda-1)\bA_0(\sigma)+c_0\,.
\end{align}
\end{subequations}
Except for the operators $E_{i0,00}$ given by Eq.~\eqref{eq:2op+0}, all
the operators in Eq.~\eqref{eq:gqsop1} leave invariant both the subspaces
in Eq.~\eqref{eq:mtypC} for \emph{arbitrary} positive-integer values of
$\cN_1$ and $\cN_2$. Therefore, the operator $\tcH$ is solvable if all the
coefficients $A_{i0,00}$ vanish, or equivalently, if
\begin{align}
\label{eq:solcn}
\bA_0 (\sigma)=0\,.
\end{align}

\section{Extraction of Schr\"odinger Operators}
\label{sec:reduc}

In the preceding section, we have constructed the most general
quasi-solvable second-order operator $\tH$ preserving the type C space
\eqref{eq:mtypC}. By applying a similarity transformation on $\tH$, we may
obtain a family of quasi-solvable operators. However, second-order
linear differential operators of several variables are in general not
equivalent to Schr\"odinger operators. This fact is one of the most
obstacles in constructing a quasi-solvable \emph{quantum} many-body systems.
Recently in Refs.~\cite{Ta04,Ta03c}, it was shown that the amount of the
difficulty can be significantly reduced by considering the underlying
symmetry of the invariant space of quasi-solvable operators. In this section,
we will show that the symmetry consideration is indeed quite efficient
in our case too.

To begin with, let us briefly review the consequence in the case of the
type A space \eqref{eq:mtypA} (see Ref.~\cite{Ta03c} for the details).
The symmetry which transforms the type A space to itself is $GL(2,\bbR)$
of linear fractional transformations of $z_i$ introduced by
\begin{align}
\label{eq:fract}
z_i\mapsto \hat{z}_i=\frac{\alpha z_i+\beta}{\gamma z_i+\delta}
 \qquad (\alpha,\beta,\gamma,\delta\in\bbR\,; \Delta\equiv\alpha\delta
 -\beta\gamma\neq 0)\,.
\end{align}
Then, it is easy to see that the type A space \eqref{eq:mtypA} is invariant
under the $GL(2,\bbR)$ transformation induced by \eqref{eq:fract}:
\begin{align}
\tcV_{\cN_k}^{(A)}[z]\mapsto
 \prod_{i=1}^M (\gamma z_i+\delta)^{\cN_k-1}\tcV_{\cN_k}^{(A)}[\hat{z}]
 =\tcV_{\cN_k}^{(A)}[z]\,.
\end{align}
{}From this invariance, we can conclude that any operator $\tcH^{(A)}$
which preserves the type A space must be shape invariant under the
$GL(2,\bbR)$ transformation as follows:
\begin{align}
\label{eq:shinv}
\tcH_{\cN_k}^{(A)}[z,a_l]\mapsto&\,
 \prod_{i=1}^M (\gamma z_i+\delta)^{\cN_k-1}\tcH_{\cN_k}^{(A)}
 [\hat{z},a_l]\prod_{i=1}^M (\gamma z_i+\delta)^{-(\cN_k-1)}\notag\\
&\,=\tcH_{\cN_k}^{(A)}[z,\hat{a}_l]\,,
\end{align}
where $a_l$ stands for all the free parameters involved in
$\tcH_{\cN_k}^{(A)}$. In particular, all the second-order operators
$\tH_{\cN_k}^{(A)}$ which preserve the type A space and are
equivalent to the Schr\"odinger operators as well must have the latter
shape-invariance.
{}From this requirement we can derive that $\tH_{\cN_k}^{(A)}$ must have
the following form:
\begin{align}
\label{eq:tAHam}
\tH_{\cN_k}^{(A)}=\tH_{\cN_k}^{(0)}-\tH_{\cN_k}^{(1)}
 -2c_k\,\tH_{\cN_k}^{(2)}-R_k\,,
\end{align}
where $c_k$ and $R_k$ are constants
and $\tH_{\cN_k}^{(j)}$ ($j=0,1,2$) are all shape-invariant operators
satisfying Eq.~\eqref{eq:shinv} as the following:
\begin{align}
\label{eq:tAop1}
\tH_{\cN_k}^{(0)}=&\,-\sum_{i=1}^M O_4^{(k)}(z_i)\frac{\del^2}{\del z_i^2}
 +\frac{\cN_k-2}{2}\sum_{i=1}^M O_4^{(k)\prime}(z_i)\frac{\del}{\del z_i}
 \notag\\
&\,-\frac{(\cN_k-1)(\cN_k-2)}{12}\sum_{i=1}^M O_4^{(k)\prime\prime}(z_i)\,,\\
\label{eq:tAop2}
\tH_{\cN_k}^{(1)}=&\,\sum_{i=1}^M O_2^{(k)}(z_i)\frac{\del}{\del z_i}
 -\frac{\cN_k-1}{2}\sum_{i=1}^M O_2^{(k)\prime}(z_i)\,,\\
\label{eq:tAop3}
\tH_{\cN_k}^{(2)}=&\,\sum_{i\neq j}^M \frac{O_4^{(k)}(z_i)}{z_i-z_j}
 \frac{\del}{\del z_i}-\frac{M-1}{4}\sum_{i=1}^M O_4^{(k)\prime}(z_i)
 \frac{\del}{\del z_i}\notag\\
&\,-\frac{\cN_k-1}{4}\left[\sum_{i\neq j}^M
 \frac{O_4^{(k)\prime}(z_i)}{z_i-z_j}
 -\frac{M-1}{3}\sum_{i=1}^M O_4^{(k)\prime\prime}(z_i)\right]\,.
\end{align}
The functions $O_n^{(k)}(z)$ ($n=2,4$) in Eqs.~\eqref{eq:tAop1}--%
\eqref{eq:tAop3} are polynomials of (at most) $n$th-degree and are
transformed under the $GL(2,\bbR)$ transformation according to
\begin{align}
O_n^{(k)}(z)\mapsto \Delta^{-n/2}(\gamma z+\delta)^n O_n^{(k)}(\hat{z})\,.
\end{align}

Keeping these facts in mind, let us come back to consider the type C case.
As we have mentioned previously, all the operators in Eq.~\eqref{eq:gqsop1}
and hence $\tH$ preserve each of the subspaces in Eq.~\eqref{eq:mtypC}
separately.
In other words, the second-order operator $\sigma_M^{-(k-1)\lambda}\,\tH
\sigma_M^{(k-1)\lambda}$ ($k=1,2$) leaves the type A space
$\tcV_{\cN_k}^{(A)}$ of dimension $\cN_k$ invariant.
On the other hand, the results for the type A case just summarized
tell us that these operators are equivalent to a Schr\"odinger operator
if and only if they have the form given by
Eqs.~\eqref{eq:tAHam}--\eqref{eq:tAop3}. Therefore, the general type C
operator $\tH$ \eqref{eq:gqsop2} is equivalent to a Schr\"odinger
operator of several variables if and only if the following conditions are
simultaneously satisfied (up to constants):
\begin{align}
\label{eq:tCcon}
\tH=\tH_{\cN_1}^{(A)}\,,\qquad
 \sigma_M^{-\lambda}\,\tH\sigma_M^\lambda=\tH_{\cN_2}^{(A)}\,.
\end{align}
The explicit form of the operator $\sigma_M^\lambda \,\tH_{\cN_k}^{(A)}
\sigma_M^{-\lambda}$ is calculated as
\begin{multline}
\label{eq:cftAop}
\sigma_M^\lambda \,\tH_{\cN_k}^{(A)}\sigma_M^{-\lambda}
=-\sum_{i=1}^M O_4^{(k)}(z_i)\frac{\del^2}{\del z_i^2}+\sum_{i=1}^M
 \biggl[\frac{\cN_k-2+(M-1)c_k}{2}O_4^{(k)\prime}(z_i)\\
+2\lambda O_4^{(k)}(z_i)z_i^{-1}-O_2^{(k)}(z_i)\biggr]\frac{\del}{\del z_i}
 -2c_k\sum_{i\neq j}^M \frac{O_4^{(k)}(z_i)}{z_i-z_j}\frac{\del}{\del z_i}\\
+c_k\sum_{i\neq j}^M \biggl[\frac{\cN_k-1}{2}O_4^{(k)\prime}(z_i)
 +2\lambda O_4^{(k)}(z_i)z_i^{-1}\biggr]\frac{1}{z_i-z_j}\\
-\sum_{i=1}^M \biggl[\frac{\cN_k-1}{12}\bigl(\cN_k-2+2(M-1)c_k\bigr)
 O_4^{(k)\prime\prime}(z_i)\\
+\frac{\cN_k-2+(M-1)c_k}{2}\lambda O_4^{(k)\prime}(z_i)z_i^{-1}
 +\lambda(\lambda+1)O_4^{(k)}(z_i)z_i^{-2}\\
-\frac{\cN_k-1}{2}
 O_2^{(k)\prime}(z_i)-\lambda O_2^{(k)}(z_i)z_i^{-1}\biggr]-R_k\,.
\end{multline}
It is evident that the condition \eqref{eq:tCcon} is satisfied if and only
if the coefficients of the differential operators of each order in
$\tH_{\cN_1}^{(A)}$ and $\sigma_M^\lambda\,\tH_{\cN_2}^{(A)}
\sigma_M^{-\lambda}$
coincide with each other (up to a constant for the zeroth-order). {}From
Eq.~\eqref{eq:cftAop} we thus obtain the following set of conditions:
\begin{gather}
\label{eq:tCon1}
O_4^{(1)}(z_i)=O_4^{(2)}(z_i)\equiv A(z_i)\,,\\
\label{eq:tCon2}
O_2^{(2)}(z_i)-O_2^{(1)}(z_i)=\frac{\cN_2-\cN_1}{2}A'(z_i)
 +2\lambda A(z_i)z_i^{-1}\,,\\
\sum_{i=1}^M \biggl[\frac{\cN_2-1}{2}O_2^{(2)\prime}(z_i)+\lambda O_2^{(2)}
 (z_i)z_i^{-1}-\frac{\cN_1-1}{2}O_2^{(1)\prime}(z_i)\biggr]
\hspace{0.3\textwidth}\notag\\
=\sum_{i=1}^M \biggl[\frac{\cN_2-\cN_1}{12}\bigl(\cN-3+2(M-1)c\bigr)A''(z_i)
 +\frac{\cN_2-2+(M-1)c}{2}\lambda A'(z_i)z_i^{-1}\notag\\
+\lambda(\lambda+1)A(z_i)z_i^{-2}\biggr]
 -c\sum_{i\neq j}^M \biggl[\frac{\cN_2-\cN_1}{2}A'(z_i)+2\lambda A(z_i)
 z_i^{-1}\biggr]\frac{1}{z_i-z_j}+\text{const.},
\label{eq:tCon3}
\end{gather}
where $c\equiv c_1=c_2$.
By substituting Eq.~\eqref{eq:tCon2} for $O_2^{(2)}(z_i)$ in
Eq.~\eqref{eq:tCon3}, the third condition can be expressed solely in terms
of $O_2^{(1)}(z_i)$ and $A(z_i)$:
\begin{multline}
\label{eq:tCon4}
\sum_{i=1}^M \biggl[\frac{\cN_2-\cN_1}{2}O_2^{(1)\prime}(z_i)
 +\lambda O_2^{(1)}(z_i)z_i^{-1}\biggr]\\
=\sum_{i=1}^M \biggl[\frac{\cN_2-\cN_1}{12}\bigl(\cN_1-2\cN_2+2(M-1)c\bigr)
 A''(z_i)\hspace{0.2\textwidth}\\
+\frac{\cN_1-2\cN_2+(M-1)c}{2}\lambda A'(z_i)z_i^{-1}
 +\lambda(\cN_2-\lambda)A(z_i)z_i^{-2}\biggr]\\
-c\sum_{i\neq j}^M \biggl[\frac{\cN_2-\cN_1}{2}A'(z_i)+2\lambda A(z_i)
 z_i^{-1}\biggr]\frac{1}{z_i-z_j}+\text{const.}
\end{multline}
On the other hand, we note that the l.h.s. of Eq.~\eqref{eq:tCon2} is
a polynomial of at most second-degree.
Hence in the polynomial $A(z)$ of at most fourth-degree defined by
Eq.~\eqref{eq:tCon1}, the coefficients of the fourth- and zeroth-degree
must vanish in order that Eq.~\eqref{eq:tCon2} is met. We can thus put
\begin{align}
\label{eq:AandO}
A(z)=a_3 z^3+a_2 z^2+a_1 z\,,\qquad
 O_2^{(1)}(z)=b_2^{(1)}z^2+b_1^{(1)}z+b_0^{(1)}\,.
\end{align}
Substituting the ansatz \eqref{eq:AandO} in Eq.~\eqref{eq:tCon3},
we find the third condition \eqref{eq:tCon3} is fulfilled if and only if
\begin{subequations}
\label{eqs:relco}
\begin{align}
b_2^{(1)}&=\frac{\cN_1-2\cN_2-2\lambda-(M-1)c}{2}\,a_3\,,\\
b_0^{(1)}&=\frac{\cN_1-2\lambda+(M-1)c}{2}\,a_1\,.
\end{align}
\end{subequations}
Finally from Eqs.~\eqref{eq:tCcon}, \eqref{eq:cftAop}, \eqref{eq:AandO},
and \eqref{eqs:relco}, we find that the second-order operators
$\tH$ which preserve the type C space \eqref{eq:mtypC} and are
equivalent to a Schr\"odinger operator as well must have the following form:
\begin{align}
\label{eq:gCHam}
\tH=-\sum_{i=1}^M A(z_i)\frac{\del^2}{\del z_i^2}-\sum_{i=1}^M B(z_i)
 \frac{\del}{\del z_i}-2c\sum_{i\neq j}^M \frac{A(z_i)}{z_i-z_j}
 \frac{\del}{\del z_i}-\bC\bigl(\sigma(z)\bigr)\,,
\end{align}
where $A(z)$ and $B(z)$ are given by
\begin{align}
\label{eq:funAz}
A(z)=&\,a_3 z^3+a_2 z^2+a_1 z\,,\\
B(z)=&\,-\bigl(\cN+\lambda-3+2(M-1)c\bigr)a_3 z^2\notag\\
 &\,+\bigl(b_1^{(1)}-[\cN_1-2+(M-1)c]a_2\bigr)z-(\lambda-1)a_1\,.
\label{eq:funBz}
\end{align}
Furthermore, the function $\bC$ in Eq.~\eqref{eq:gCHam} is calculated as
\begin{align}
\label{eq:funbC}
\bC\bigl(\sigma(z)\bigr)=&\,\frac{\cN_1-1}{12}\bigl(\cN_1-2+2(M-1)c\bigr)
 \sum_{i=1}^M A''(z_i)\notag\\
&\,-\frac{\cN_1-1}{2}\sum_{i=1}^M O_2^{(1)}(z_i)-\frac{\cN_1-1}{2}c
 \sum_{i\neq j}^M \frac{A'(z_i)}{z_i-z_j}+R_1\notag\\
=&\,(\cN_1-1)(\cN_2+\lambda-1)a_3\sum_{i=1}^M z_i+c_0\,,
\end{align}
where the constants $R_1$ and $c_0$ are related with each other so that $c_0$
in Eq.~\eqref{eq:funbC} exactly coincides with that in Eq.~\eqref{eq:polbC}
through the following relation:
\begin{align}
c_0=\frac{\cN_1-1}{6}\bigl(\cN_1-2-(M-1)c\bigr)Ma_2
 -\frac{\cN_1-1}{2}M b_1^{(1)}+R_1\,.
\end{align}
In fact, it is easily shown that the gauged Hamiltonian \eqref{eq:gCHam}
can be cast into a Schr\"odinger operator by a gauge transformation
\begin{align}
H&=\ee^{-\cW}\tH\ee^\cW\notag\\
&=-\frac12 \sum_{i=1}^M \frac{\del^2}{\del q_i^2}+\frac12 \sum_{i=1}^M
 \left[\left(\frac{\del\cW}{\del q_i}\right)^2-\frac{\del^2 \cW}{\del q_i^2}
 \right]-\bC\bigl(\sigma(z)\bigr)\,,
\label{eq:tCHam}
\end{align}
if the gauge potential $\cW$ is chosen as
\begin{align}
\label{eq:tCgau1}
\cW=-\sum_{i=1}^M \int\dd z_i\,\frac{B(z_i)}{2A(z_i)}+\frac14 \sum_{i=1}^M
 \ln \bigl|A(z_i)\bigr|-c\sum_{i<j}^M \ln |z_i-z_j|\,,
\end{align}
and the function $z(q)$ which determines the change of variables satisfies
\begin{align}
\label{eq:zqdef}
z'(q)^2=2A\bigl(z(q)\bigr)=2\bigl(a_3 z(q)^3+a_2 z(q)^2+a_1 z(q)\bigr)\,.
\end{align}
To make the correspondence between the results here and those in
Ref.~\cite{GT04a} more transparent, let us decompose $B(z)$ as
\begin{align}
B(z)=Q(z)-\frac{\cN-2+2(M-1)c}{2}A'(z)\,,
\end{align}
so that the coefficients of the second- and zeroth-degree in $Q(z)$
coincide with each other:
\begin{align}
\label{eq:defQz}
Q(z)=\frac{\cN-2\lambda+2(M-1)c}{2}A(z)z^{-1}-2\beta z\,.
\end{align}
The parameter $\beta$ in Eq.~\eqref{eq:defQz} is given by
\begin{align}
2\beta=\frac{\cN_1-\cN_2-2\lambda}{2}\,a_2-b_1^{(1)}\,.
\end{align}
Then, if we put $b_1$ as the coefficient of the first-degree in $Q(z)$
and take $M=1$, we can easily check that Eqs.~\eqref{eq:gCHam}--%
\eqref{eq:defQz} completely reduce to the corresponding quantities in
Ref.~\cite{GT04a} as they should.

It is apparent from the construction that the Hamiltonian \eqref{eq:tCHam}
preserves the space $\cV_{\cN_1\!,\cN_2;M}^{(C)}$ defined by
\begin{align}
\label{eq:invsp}
\cV_{\cN_1\!,\cN_2;M}^{(C)}=e^{-\cW}\,\tcV_{\cN_1\!,\cN_2;M}^{(C)}\,.
\end{align}
Hence, the Hamiltonian $H$ can be (locally) diagonalized in this finite
dimensional space \eqref{eq:invsp}. We shall thus call
the space \eqref{eq:invsp} the \emph{solvable sector} of $H$.
To obtain the explicit form of $\cV_{\cN_1\!,\cN_2;M}^{(C)}$,
we first arrange the gauge potential \eqref{eq:tCgau1} in a more tractable
form:
\begin{align}
\cW=&\,-\frac{\alpha^-}2 \sum_{i=1}^M \ln |z_i|-\frac{\cN-1+2(M-1)c}{4}
 \sum_{i=1}^M \ln\left|\frac{z_i}{A(z_i)}\right|\notag\\
&\,+\beta\sum_{i=1}^M \int\dd z_i\,\frac{z_i}{A(z_i)}-c\sum_{i<j}^M
 \ln |z_i-z_j|\,,
\label{eq:tCgau2}
\end{align}
where a new parameter $\alpha^-$ is introduced by
\begin{align}
\alpha^- =\frac{1-2\lambda}{2}\,.
\end{align}
Then, from Eqs.~\eqref{eq:mtypC}, \eqref{eq:invsp}, and \eqref{eq:tCgau2},
we have
\begin{align}
\cV_{\cN_1\!,\cN_2;M}^{(C)}=&\,\ee^{-\cU}\sigma_M(z)^{\frac12 \alpha^-}\,
 \tcV_{\cN_1;M}^{(A)}[z]\oplus\ee^{-\cU}\sigma_M(z)^{\frac12(1-\alpha^-)}\,
 \tcV_{\cN_2;M}^{(A)}[z]\Bigr|_{z=z(q)}\notag\\
\equiv&\,\cV_{\cN_1;M}^{(1)}\oplus\cV_{\cN_2;M}^{(2)}\,,
\label{eq:solsc}
\end{align}
where the new gauge factor $\ee^{-\cU}$ is defined by
\begin{align}
\ee^{-\cU}=&\,\prod_{i<j}^M |z_i-z_j|^c \,\prod_{i=1}^M \left|
 \frac{z_i}{A(z_i)}\right|^{\frac14\left(\cN-1+2(M-1)c\right)}\notag\\
&\,\times\exp\left(-\beta\sum_{i=1}^M
 \int\dd z_i\,\frac{z_i}{A(z_i)}\right)\,.
\end{align}

\section{Gauged Hamiltonians in the general type C operators}
\label{sec:inver}

In the preceding section, we have constructed the type C gauged Hamiltonian
$\tH$ \eqref{eq:gCHam} almost independently of the explicit form of the most
general type C operator $\tcH$ given by Eq.~\eqref{eq:gqsop2}. In fact,
all that we have employed in the previous construction is the known results
for the many-body type A gauged Hamiltonians and the fact that $\tcH$
preserves the two type A spaces in Eq.~\eqref{eq:mtypC} separately.
In this section, we will thus show how
the gauged Hamiltonian $\tH$ is embedded in the general type C
operator $\tcH$. For this purpose, we shall express $\tH$ in terms
of $\sigma_k$. The necessary formulas are summarized in Appendix.

First from Eqs.~\eqref{eq:polbC} and \eqref{eq:funbC}, we easily see that
$\bA_0$ having the form \eqref{eq:polA0} must be
\begin{align}
\label{eq:expA0}
\bA_0(\sigma)=a_3 \sigma_1\,.
\end{align}
With the aid of Eqs.~\eqref{eqs:form1} and \eqref{eq:expA0}, the
second-order operator in Eq.~\eqref{eq:gCHam} is expressed in terms of
$\sigma_k$ as
\begin{align}
\label{eq:2ndCH}
-\sum_{i=1}^M A(z_i)\frac{\del^2}{\del z_i^2}=-\sum_{k,l=1}^M
 \bigl[ \bA_0(\sigma)\sigma_k \sigma_l +\cA_{kl}(\sigma)\bigr]
 \frac{\del^2}{\del\sigma_k \del\sigma_l}\,,
\end{align}
where $\cA_{kl}$ are polynomials of second-degree as the following:
\begin{align}
\cA_{kl}(\sigma)=&\,-a_3 \biggl[\,\sum_{m=0}^1 (k-m+1)\sigma_{k-m+1}
 \sigma_{l+m}\notag\\
&\,\qquad+\sum_{m=2}^{k+1}(k-l-2m+1)\sigma_{k-m+1}\sigma_{l+m}\biggr]\notag\\
&\,+a_2 \biggl[k\,\sigma_{k}\sigma_{l}+\sum_{m=1}^{k}(k-l-2m)
 \sigma_{k-m}\sigma_{l+m}\biggr]\notag\\
&\,-a_1 \sum_{m=0}^{k-1}(k-l-2m-1)\sigma_{k-m-1}\sigma_{l+m}\,.
\label{eq:cAkls}
\end{align}
Comparing Eq.~\eqref{eq:2ndCH} with the second-order operators in
Eq.~\eqref{eq:gqsop2}, we immediately obtain
\begin{align}
\label{eq:expAkl}
\bA_{\bk\bl}(\sigma)&=\cA_{\bk\bl}(\sigma)+\cA_{\bl\bk}(\sigma)
 \qquad (\bk>\bl)\,,\\
\label{eq:expAkk}
\bA_{\bk\bk}(\sigma)&=\cA_{\bk\bk}(\sigma)\,,\\
\label{eq:cnAMk}
\bA_{M\!\bk}(\sigma)\sigma_M &=\cA_{\bk M}(\sigma)+\cA_{M\!\bk}(\sigma)\,,\\
\bA_{M\!M}(\sigma)\sigma_M &=\cA_{M\!M}(\sigma)\,.
\end{align}
The r.h.s. of Eqs.~\eqref{eq:expAkl} and \eqref{eq:expAkk} indeed have
the same forms as Eqs.~\eqref{eq:polAkl} and thus completely determine
all the polynomials $\bA_{\bk\bl}$ ($\bk\geq\bl$).
To factor out $\sigma_M$ in the r.h.s. of Eq.~\eqref{eq:cnAMk},
the following formula is useful:
\begin{align}
\label{eq:smsig}
\sum_{m=0}^{M-k}(M-k-2m-1)\sigma_{M-m-1}\sigma_{k+m}
 =(k-M-1)\sigma_{k-1}\sigma_{M}\,.
\end{align}
{}From Eqs.~\eqref{eq:cAkls}--\eqref{eq:smsig}, the first-degree polynomials
$\bA_{M\!k}$ of the form \eqref{eq:polAMk} read
\begin{align}
\label{eq:expAMk}
\bA_{M\!\bk}(\sigma)&=-2(\bk+1)a_3 \sigma_{\bk+1}+2\bk a_2 \sigma_{\bk}
 -2(\bk-M-1)a_1 \sigma_{\bk-1}\,,\\
\label{eq:expAMM}
\bA_{M\!M}(\sigma)&=Ma_2 \sigma_M +a_1 \sigma_{M-1}\,.
\end{align}
Similarly, with the aid of the formulas \eqref{eqs:form2}, \eqref{eqs:form3}
and Eq.~\eqref{eq:expA0} the first-order operators in Eq.~\eqref{eq:gCHam}
are expressed in terms of $\sigma_k$ as
\begin{align}
-\sum_{i=1}^M B(z_i)\frac{\del}{\del z_i}&-2c\sum_{i\neq j}^M
 \frac{A(z_i)}{z_i-z_j}\frac{\del}{\del z_i}\notag\\
&=\sum_{k=1}^M \bigl[ (\cN+\lambda-3)\bA_0 (\sigma)-\cB_k (\sigma)
 \bigr]\frac{\del}{\del\sigma_k}\,,
\end{align}
where $\cB_k$ are polynomials of first-degree as the following:
\begin{multline}
\cB_k (\sigma)=(k+1)(\cN+\lambda-3+kc)a_3 \sigma_{k+1}\\
-\frac{k}2 \Bigl([\cN+2\lambda-4+2(k-M)c]a_2+4\beta\Bigr)\sigma_k\\
+(k-M-1)\bigl(\lambda-1+(k-M)c\bigr)a_1 \sigma_{k-1}\,.
\end{multline}
In particular, $\cB_M$ can be arranged with the use of
Eq.~\eqref{eq:expAMM} as
\begin{align}
\cB_M(\sigma)=-(\lambda-1)\bA_{M\!M}(\sigma)-\frac{M}2 \bigl(
 (\cN-2)a_2 +4\beta\bigr)\sigma_M\,.
\end{align}
It is now apparent that $\cB_k$ are all in agreement with the form of
$\bB_k$ given by Eqs.~\eqref{eq:polBk} and \eqref{eq:polBM} and thus
\begin{align}
\label{eq:expBk}
\bB_k (\sigma)=\cB_k (\sigma)\,.
\end{align}
It should be pointed out that after determining all the parameters $A_{ik,jl}$
from Eqs.~\eqref{eq:expAkl}, \eqref{eq:expAkk}, \eqref{eq:expAMk}, and
\eqref{eq:expAMM} we can always adjust the parameters $B_{i\bk}$ and
$B_{0\bk}$ in Eq.~\eqref{eq:polBk} and $B_{M\!M}$ in Eq.~\eqref{eq:polBM}
so that Eq.~\eqref{eq:expBk} holds. Therefore, the type C gauged Hamiltonian
\eqref{eq:gCHam} in fact constitutes a subclass of the general type C operator
\eqref{eq:gqsop2}.

{}From Eq.~\eqref{eq:expA0} and the discussion following Eq.~\eqref{eq:polbC},
it is apparent that the gauged Hamiltonian $\tH$ is solvable if
\begin{align}
\label{eq:solcn2}
a_3 =0\,.
\end{align}
In this case, the polynomial $\bC$ given by Eq.~\eqref{eq:funbC} is
a constant and hence the many-body Hamiltonian \eqref{eq:tCHam} is of
the supersymmetric form up to this irrelevant constant. The gauge potential
$\cW$ can be regarded as a superpotential.

\section{Classification of the Models}
\label{sec:class}

We shall now explicitly compute the type C models. The potential term in
Eq.~\eqref{eq:tCHam} is explicitly calculated in terms of $z_i$ from
Eqs.~\eqref{eq:funAz}--\eqref{eq:tCgau1} as
\begin{align}
\label{eq:tCpot}
V=&\,\sum_{i=1}^M \frac{1}{4A(z_i)}\biggl[\frac{\cN-1+2(M-1)c}{2}A'(z_i)
 -Q(z_i)\biggr]\notag\\
&\,\times\biggl[\frac{\cN+1+2(M-1)c}{2}A'(z_i)-Q(z_i)\biggr]
 -a_3(M,\cN_i,\lambda)\sum_{i=1}^M z_i\notag\\
&\,+c(c-1)\sum_{i<j}^M \frac{A(z_i)+A(z_j)}{(z_i-z_j)^2}+V_0\,,
\end{align}
where, and in what follows, $V_0$ denotes an arbitrary constant, and
the coupling constant $a_3(M,\cN_i,\lambda)$ is given by
\begin{align}
a_3(M,\cN_i,\lambda)=&\,\frac12 \bigl[ 2\cN_1(\cN_2+\lambda)
 +2(\cN+\lambda)(M-1)c\notag\\
&\,+2M(M-1)c^2-2(M-1)c-1\bigr]a_3\,.
\end{align}
{}From Eq.~\eqref{eq:zqdef}, the change of variable is determined by
the elliptic integral:
\begin{align}
\label{eq:ellint}
\pm(q-q_0)=\int\frac{\dd z}{\sqrt{2(a_3 z^3+a_2 z^2+a_1 z)}}\,.
\end{align}
In contrast to the systems constructed from the $\Lsl(M+1)$ generators,
our present models do not have full $GL(2,\bbR)$ invariance; the conditions
\eqref{eq:tCon2} and \eqref{eq:tCon3} apparently break the $GL(2,\bbR)$
symmetry. However, there remains residual symmetry unbroken, namely, the
invariance under scale transformations, $\beta=\gamma=0$ in
Eq.~\eqref{eq:fract},
and under special projective transformations, $\alpha=\delta=0$ and
$\beta=\gamma(\neq 0)$ in Eq.~\eqref{eq:fract}. The latter invariance
is as it should be since we have made the generalization so that
Eq.~\eqref{eq:minvC} holds. Due to the invariance, we can classify
the models completely the same way as in the single-variable case presented
in Ref.~\cite{GT04a}. This is another virtue of our choice of the space
\eqref{eq:mtypC} based on the invariance criterion.
It is readily shown that $A(z)$ can be cast into one of the canonical forms
listed in Table~\ref{tb:table1} by the combination of the aforementioned scale
and projective transformations. For the detailed treatment of the elliptic
integral \eqref{eq:ellint} in each case, see Ref.~\cite{GT04a}.
\begin{table}
\caption{Canonical forms for the polynomial $A(z)$, \eqref{eq:funAz}.}
\label{tb:table1}
\vspace*{.6cm}
\begin{center}
\begin{tabular}{ll}
\hline
Case & Canonical Form\\
\hline
1 & $2z$\\
2 & $\pm 2\nu z^2$\\
3 & $\pm 2\nu z(1-z)(1-m z)$\\ 
4 & $2\nu z(1-z)(m'+m z)$\\
5 & $\frac12\nu z\bigl(z^2+2(1-2m)z+1\bigr)$\\
\hline 
\end{tabular}
\end{center}
\vspace*{.6cm}
In this Table, $\nu>0$, $0\leq m\leq 1$ and $m'=1-m$ ($m\neq1$ in Case 3 and
$m\neq0,1$ in Case 4 to avoid duplications).
\noindent
\end{table}

Furthermore, we note that from Eq.~\eqref{eq:zqdef}, a rescaling of
the coefficients $a_i$, $\beta$, $c_0$ by an overall constant factor
$\nu>0$ has the following effect on the change of variable $z(q)$:
\begin{align}
\label{eq:zscale}
z(q\,;\nu a_i,\nu\beta,\nu c_0)=z(\rnu\,q\,;a_i,\beta,c_0)\,.
\end{align}
{}From this equation and Eqs.~\eqref{eq:funAz}--\eqref{eq:funbC},
\eqref{eq:tCHam}, and \eqref{eq:tCgau1} we easily
obtain the identities
\begin{subequations}
\label{eq:EFWV}
\begin{align}
\label{eq:cWscale}
\cW(q\,;\nu a_i,\nu\beta,\nu c_0)&=\cW(\rnu\,q\,;a_i,\beta,c_0)\\
\label{eq:Vscale}
V(q\,;\nu a_i,\nu\beta,\nu c_0) &=
 \nu\,V(\rnu\,q\,;a_i,\beta,c_0)\,.
\end{align}
\end{subequations}
We shall therefore set $\nu=1$ in the canonical forms in Cases 2--5,
the models corresponding to an arbitrary value of $\nu$ following easily from
Eqs.~\eqref{eq:zscale} and \eqref{eq:EFWV}. It should also be obvious from
Eq.~\eqref{eq:ellint} that the change of variable $z(q)$, and hence the
potential $V$ determining each model, are defined up to the
transformation $q\mapsto\pm(q-q_0)$, where $q_0\in\bbR$ is a constant.
In the following, another new parameter $\alpha^+$ is introduced by
\begin{align}
\alpha^+ =\cN_2-\cN_1+\lambda+\frac12\,.
\end{align}
The condition for solvability \eqref{eq:solcn2} implies that the models
corresponding to the first and second canonical forms in Table~\ref{tb:table1},
and to the third one with $m=0$, are not only quasi-solvable but also solvable.

As we will see below, the potentials $V$ in all the cases but the irrelevant
Case 2 have the singularities at $q_i=0$ and at integer multiples of a real
number $\Omega$, $q_i=n\Omega$ ($n\in\bbZ$), where $\Omega$ is a submultiple
of the real period of the potential $V$ or $\Omega=\infty$ when the potential
is non-periodic, as well as $q_i=\pm q_j$ for $i\neq j$. Hence,
the Hamiltonians are naturally defined on
\begin{align}
0<q_M <\dots<q_1 <\Omega\,.
\end{align}
The potential form of the one-body part in each case is completely
identical with that of one-body type C $\cN$-fold supersymmetric potential
$V^-$ in Ref.~\cite{GT04a}. Thus, the normalizability conditions coming
from the one-body part, which restrict the values of the parameters
$\alpha^\pm$ and $\beta^\pm$ in each case, are completely the same as
those shown in this reference. Only an additional condition arises from
the two-body part. The finiteness of the $L^2$ norm of the two-body
wave function in the solvable sector $\cV_{\cN_1\!,\cN_2;M}^{(C)}$ leads
$c>-1/2$, where $c$ denotes the coupling constant of the two-body
interaction appeared in the last line of Eq.~\eqref{eq:tCpot}.

\subsection{Case 1: $A(z)=2z$}

\emph{Change of variable}:\quad $z=q^2$.\\

\emph{Potential}:
\begin{align}
\label{eq:pots1}
V=&\,\frac{\beta^2}{2}\sum_{i=1}^M q_i^2+\frac{\alpha^-(\alpha^--1)}{2}
 \sum_{i=1}^M \frac{1}{q_i^2}\notag\\
&\,+c(c-1)\sum_{i<j}^M \left[\frac{1}{(q_i-q_j)^2}+\frac{1}{(q_i+q_j)^2}
 \right]+V_0\,.
\end{align}
\emph{Solvable sectors}:
\begin{subequations}
\label{eq:ssec1}
\begin{align}
\label{eq:ssec1a}
\cV_{\cN_1;M}^{(1)}&=\prod_{i<j}^M \left|q_i^2-q_j^2\right|^c
 \biggl(\,\prod_{i=1}^M
 q_i^{\alpha^-}\biggr)\exp\biggl(-\frac{\beta}{2} \sum_{i=1}^M q_i^2\biggr)\,
 \tcV_{\cN_1;M}^{(A)}\bigl[q^2\bigr]\,,\\
\label{eq:ssec1b}
\cV_{\cN_2;M}^{(2)}&=\prod_{i<j}^M \left|q_i^2-q_j^2\right|^c
 \biggl(\,\prod_{i=1}^M
 q_i^{1-\alpha^-}\biggr)\exp\biggl(-\frac{\beta}{2} \sum_{i=1}^M q_i^2\biggr)
 \,\tcV_{\cN_2;M}^{(A)}\bigl[q^2\bigr]\,,
\end{align}
\end{subequations}
where $\cV_{\cN_k;M}^{(k)}$ ($k=1,2$) are defined in Eq.~\eqref{eq:solsc}.
This case corresponds to the rational $BC_M$ type Calogero-Sutherland
model~\cite{OP83}. The one-body part of the potential is only singular at
$q_i=0$ and hence a natural choice is $\Omega=\infty$.

\subsection{Case 2a: $A(z)=2 z^2$}

\emph{Change of variable}:\quad $z=\ee^{2q}$.\\

\emph{Potential}:
\begin{align}
\label{eq:pots2}
V = c(c-1)\sum_{i<j}^M \frac{1}{\sinh^2 (q_i-q_j)}+V_0\,.
\end{align}
\emph{Solvable sectors}:
\begin{subequations}
\label{eq:ssec2}
\begin{align}
\label{eq:ssec2a}
\cV_{\cN_1;M}^{(1)}&=\prod_{i<j}^M \left|\ee^{2 q_i}-\ee^{2 q_j}\right|^c
 \biggl(\,\prod_{i=1}^M \ee^{(\alpha^- -\bar{\beta}) q_i}\biggr)
 \tcV_{\cN_1;M}^{(A)}\bigl[\ee^{2q}\bigr]\,,\\
\label{eq:ssec2b}
\cV_{\cN_2;M}^{(2)}&=\prod_{i<j}^M \left|\ee^{2 q_i}-\ee^{2 q_j}\right|^c
 \biggl(\,\prod_{i=1}^M \ee^{(1-\alpha^- -\bar{\beta}) q_i}\biggr)
 \tcV_{\cN_2;M}^{(A)}\bigl[\ee^{2q}\bigr]\,.
\end{align}
\end{subequations}
\emph{Parameter}:
\begin{align}
\label{eq:para2}
\bar{\beta}=\beta+\frac{\cN-1+2(M-1)c}{2}\,.
\end{align}
We note that the potential \eqref{eq:pots2} depends on neither the parameter
$\alpha^-$ nor $\bar{\beta}$ appeared in the solvable sectors \eqref{eq:ssec2}.
The dependence on these parameters only comes from the middle part of the
each solvable sector. These parts are nothing but the wave functions
corresponding to the free motion of the center-of-mass due to the translational
invariance of the potential \eqref{eq:pots2} and thus irrelevant.
The remaining factors in Eqs.~\eqref{eq:ssec2a} and \eqref{eq:ssec2b}
coincide with each other and the dependence on $\cN_k$ is not relevant
since this case is solvable. Therefore, Case 2 only reproduces the well-known
result of the (hyperbolic) $A_{M-1}$ Sutherland model~\cite{Su71}.

\subsection{Case 2b: $A(z)=-2 z^2$}
The formula for the potential for this case can be
easily deduced from those of the preceding one by applying
Eqs.~\eqref{eq:EFWV} with $\nu=-1$. Apparently, this case only shows
the usual solvability of the trigonometric $A_{M-1}$ Sutherland
model.

\subsection{Case 3a: $A(z)=2z(1-z)(1-mz)$}

\emph{Change of variable}:\quad $z=\sn^2 q$.\\

Here, and in the following cases, the Jacobi elliptic functions have
modulus $k=\sqrt m$.\\

\emph{Potential}:
\begin{multline}
\label{eq:pots3}
V = \frac{m\alpha^+(\alpha^+ -1)}{2}\sum_{i=1}^M \sn^2 q_i
 +\frac{\alpha^-(\alpha^- -1)}{2}\sum_{i=1}^M \frac{1}{\sn^2 q_i}\\
+\frac{m'\beta^-(\beta^- -1)}{2}\sum_{i=1}^M \frac{1}{\cn^2 q_i}
 -\frac{m'\beta^+(\beta^+ -1)}{2}\sum_{i=1}^M \frac{1}{\dn^2 q_i}\\
+c(c-1)\sum_{i<j}^M \left[\frac{1}{\sn^2 (q_i-q_j)}
 +\frac{1}{\sn^2 (q_i+q_j)}\right]+V_0\,.
\end{multline}
\emph{Parameters}:
\begin{align}
\label{eq:bm3}
\beta^\pm=\mp\frac{\beta}{m'}-\frac12\bigl(\cN-1+2(M-1)c\bigr)\,.
\end{align}
\emph{Solvable sectors}:
\begin{subequations}
\label{eq:ssec3}
\begin{align}
\cV_{\cN_{1};M}^{(1)}=&\,\prod_{i<j}^M \left|\sn^2 q_i-\sn^2 q_j\right|^c
 \notag\\
&\,\times\prod_{i=1}^M \bigl[(\sn q_i)^{\alpha^-}(\cn q_i)^{\beta^-}
 (\dn q_i)^{\beta^+}\bigr]
 \,\,\tcV_{\cN_1;M}^{(A)}\bigl[\sn^2 q\bigr]\,,\\
\cV_{\cN_{2};M}^{(2)}=&\,\prod_{i<j}^M \left|\sn^2 q_i-\sn^2 q_j\right|^c
 \notag\\
&\,\times\prod_{i=1}^M \bigl[(\sn q_i)^{1-\alpha^-}(\cn q_i)^{\beta^-}
 (\dn q_i)^{\beta^+}\bigr]
 \,\,\tcV_{\cN_2;M}^{(A)}\bigl[\sn^2 q\bigr]\,.
\end{align}
\end{subequations}
We note that from the formula
\begin{align}
\label{eq:dblfm}
\frac{m'}{\cn^2 q}=-m\sn^2 q-\frac1{\sn^2 q}+\frac4{\sn^2 2q}
 +\frac{m'}{\dn^2 q}-(1+m)\,,
\end{align}
we can regard the above model \eqref{eq:pots3} as a deformation of
an Olshanetsky--Perelomov type potential associated with the $BC_M$ root
system~\cite{OP83}.
In contrast to the ordinary elliptic integrable models, the potential
function here is given by the Jacobian function instead of the Weierstrass
$\wp$ function. We thus call the potential \eqref{eq:pots3}
\emph{Jacobi-elliptic $BC_M$ Inozemtsev model}.
The one-body part of the potential have real singularities at the points
$q_i=nK(m)$ ($n\in\bbZ$) where $K(m)$ is the complete elliptic integral
of the first kind defined by
\begin{align}
\label{eq:ceii}
  K(m)\equiv\int_0^{\pi/2}\frac{dt}{\sqrt{1-m\sin^2 t}}\,.
\end{align}
Hence a natural choice is $\Omega=K(m)$.

As previously mentioned, when $m=0$ the polynomial
$A(z)$ is of second degree and hence the Hamiltonian $H$ is solvable.
The formulas for the potential and the solvable sectors
are obtained from Eqs.~\eqref{eq:pots3}--\eqref{eq:ssec3} by setting $m=0$,
$m'=1$, and $(\sn q,\cn q,\dn q)=(\sin q,\cos q,1)$. With the aid of the
formula \eqref{eq:dblfm}, we immediately see that the potential
\eqref{eq:pots3} in this case is nothing but the trigonometric $BC_M$
Calogero--Sutherland model.

\subsection{Case 3b: $A(z)=-2z(1-z)(1-m\,z)$}

As in Case 2b, the formulas for this case can follow from those of the
preceding one using Eqs.~\eqref{eq:EFWV} with $\nu=-1$.
The following well-known identities \cite{GR00} may be of help in this case:
\[
  \sn(\ii q\,;m) = \ii\,\frac{\sn(q\,;m')}{\cn(q\,;m')}\,,\quad
  \cn(\ii q\,;m) = \frac1{\cn(q\,;m')}\,,\quad
  \dn(\ii q\,;m) = \frac{\dn(q\,;m')}{\cn(q\,;m')}\,.
\]
The real singularities of the one-body part of the potential are now at
$q_i=nK(m')$ ($n\in\bbZ$) and thus we naturally choose as $\Omega=K(m')$.
The value $m=0$ yields again solvable model, whose potentials
and solvable sectors follow from Eqs.~\eqref{eq:pots3}--\eqref{eq:ssec3} by
setting $m=0$, $m'=1$, and $(\sn\ii q,\cn\ii q,\dn\ii q)=(\ii\sinh q,
\cosh q,1)$. Clearly, the potential in this case corresponds to the
hyperbolic $BC_M$ Calogero-Sutherland model.

\subsection{Case 4: $A(z)=2z(1-z)(m'+mz)$}

\emph{Change of variable}:\quad $z=\cn^2 q$.\\

\emph{Potential}:
\begin{multline}
\label{eq:pots4}
V = \frac {m\alpha^+(\alpha^+ -1)}{2}\sum_{i=1}^M \sn^2 q_i
 +\frac{m'\alpha^-(\alpha^- -1)}{2}\sum_{i=1}^M \frac{1}{\cn^2 q_i}\\
+\frac{\beta^-(\beta^- -1)}{2}\sum_{i=1}^M \frac{1}{\sn^2 q_i}
 -\frac{m'\beta^+(\beta^+ -1)}{2}\sum_{i=1}^M \frac{1}{\dn^2 q_i}\\
+c(c-1)\sum_{i<j}^M \left[\frac{1}{\sn^2(q_i-q_j)}
 +\frac{1}{\sn^2(q_i+q_j)}\right]+ V_0\,.
\end{multline}
\emph{Parameters}:
\begin{align}
\label{eq:bm4}
\beta^\pm=\mp\beta-\frac12 \bigl(\cN-1+2(M-1)c\bigr)\,.
\end{align}
\emph{Solvable sectors}:
\begin{subequations}
\label{eq:ssec4}
\begin{align}
\cV_{\cN_1;M}^{(1)}=&\,\prod_{i=1}^M \left|\sn^2 q_i-\sn^2 q_j\right|^c\notag\\
&\,\times\prod_{i=1}^M \bigl[(\cn q_i)^{\alpha^-}(\sn q_i)^{\beta^-}
 (\dn q_i)^{\beta^+}\bigr]\,\,\tcV_{\cN_1;M}^{(A)}\bigl[\cn^2 q\bigr]\,,\\
\cV_{\cN_2;M}^{(2)}=&\,\prod_{i=1}^M \left|\sn^2 q_i-\sn^2 q_j\right|^c\notag\\
&\,\times\prod_{i=1}^M \bigl[(\cn q_i)^{1-\alpha^-}(\sn q_i)^{\beta^-}
 (\dn q_i)^{\beta^+}\bigr]\,\,\tcV_{\cN_2;M}^{(A)}\bigl[\cn^2 q\bigr]\,.
\end{align}
\end{subequations}
We again obtain a Jacobi-elliptic $BC_M$ Inozemtsev model.
The location of the real singularities are completely the same as that in
Case 3a.

\subsection{Case 5: $A(z)= z\bigl(z^2+2(1-2m)z+1\bigr)/2$}

\emph{Change of variable}:\quad $z=(1+\cn q)/(1-\cn q)$.\\

\emph{Potential}:
\begin{multline}
\label{eq:pots5}
V = \frac{\alpha^+(\alpha^+ -1)}{4}\sum_{i=1}^M \frac{1}{1-\cn q_i}
 +\frac{\alpha^-(\alpha^- -1)}{4}\sum_{i=1}^M \frac{1}{1+\cn q_i}\\
+\biggl[\frac{\beta^2}{2m}-\frac{m'}{8}\bigl(\cN-1+2(M-1)c\bigr)
 \bigl(\cN+1+2(M-1)c\bigr)\biggr]\sum_{i=1}^M \frac{1}{\dn^2 q_i}\\
+\bigl(\cN+2(M-1)c\bigr)\frac{\beta}2 \sum_{i=1}^M \frac{\cn q_i}{\dn^2 q_i}\\
+c(c-1)\sum_{i<j}^M\Biggl[\frac{(1-\cn q_i \cn q_j )
 (m'+m\cn q_i \cn q_j)}{(\cn q_i -\cn q_j)^2} \Biggr]+ V_0\,.
\end{multline}
\emph{Gauge factor}:
\begin{multline}
\label{eq:gauge5}
\ee^{-\cU^\pm}=\prod_{i<j}^M \left|\frac{1+\cn q_i}{1-\cn q_i}
 -\frac{1+\cn q_j}{1-\cn q_j}\right|^c\,
 \prod_{i=1}^M \Bigl(\frac{1-\cn q_i}{\dn q_i}
 \Bigr)^{\frac12 (\cN-1+2(M-1)c)}\\
\times\exp\left(-\frac{\beta}{k k'}\sum_{i=1}^M \arctan
 \frac{k'^2+k^2 \cn q_i}{kk'(1-\cn q_i)}\right),
\end{multline}
where $k'=\sqrt{m'}=\sqrt{1-m}$. The one-body part of the potential is
singular at integer multiples of $2K(m)$, so that a natural choice is
$\Omega=2K(m)$. This model is also regarded as a kind of Jacobi-elliptic
$BC_M$ Inozemtsev system since this case is obtained by a \emph{complex}
projective transformation from Case 3($m\neq 0$) or Case 4
(cf. Section~\ref{sec:discus}).

\section{Difficulty in Type B Generalization}
\label{sec:diffi}

Before concluding the article, we will give brief comments on the
several-variable
generalization of the type B monomial space and its difficulty.
The type B monomial space of a single variable is defined by
\begin{align}
\label{eq:stypB}
\tcV_{\cN}^{(B)}=\bigl\langle 1,z,\dots,z^{\cN-2},z^\cN \bigr\rangle\,.
\end{align}
The general quasi-solvable operators preserving the space \eqref{eq:stypB}
and the corresponding $\cN$-fold supersymmetry are investigated in
Ref.~\cite{GT03}. It is pointed out in Ref.~\cite{GT04a} that the type B
space \eqref{eq:stypB} is realized from the type C space \eqref{eq:stypC}
by the following formal substitution
\begin{align}
\label{eq:CtoB}
\cN_2=1,\qquad \lambda=\cN_1+1=\cN\,,
\end{align}
although the characteristic features of the type B and C quasi-solvable
operators are quite different from each other. Hence, a natural
generalization of the type B space is achieved by the same substitution
as Eq.~\eqref{eq:CtoB} in the type C space of several variables
\eqref{eq:mtypC}, which results in
\begin{align}
\label{eq:mtypB}
\tcV_{\cN;M}^{(B)}=\tcV_{\cN-1;M}^{(A)}
 \oplus\bigl\langle\sigma_M^\cN \bigr\rangle\,.
\end{align}
We easily see that the above space \eqref{eq:mtypB} indeed reduces to
Eq.~\eqref{eq:stypB} when $M=1$.
In contrast to the single-variable case, however, there is no second-order
operator preserving the space \eqref{eq:mtypB} which raises the degree of
the monomials by two; such an operator must delete \emph{simultaneously}
all the elements
$\sigma_1^{n_1}\dots\sigma_M^{n_M}$ with $\sum_i n_i=\cN-3$ and $\cN-2$
as well as the element $\sigma_M^\cN$, but it cannot be achieved by any
at most second-order operator.
In other words, there is no several-variable counterpart of an operator
which reduces (when $M=1$) to the degree two second-order operator $J_{++}$
preserving the single-variable type B space \eqref{eq:stypB}:
\begin{align}
J_{++}=z^2\Bigl(z\frac{\dd}{\dd z}-\cN\Bigr)
 \Bigl(z\frac{\dd}{\dd z}-\cN+3\Bigr)\,.
\end{align}
Needless to say, there are many other possibilities for the generalizations
instead of Eq.~\eqref{eq:mtypB}. However, this kind of difficulty may not be
the sole obstacle and, as far as we have investigated so far, no natural
generalization of the single-variable type B has been found.

\section{Discussion and Summary}
\label{sec:discus}

In this article, we have developed a systematic procedure to construct
a family of quasi-solvable operators in several variables leaving two type A
spaces separately and thus admitting two different polynomial type solutions.
It is evident that this family
is regarded as a subclass of the operators which leave a single type A
space \eqref{eq:mtypA} invariant and thus ensure only one polynomial type
solution. The latter operators are exactly those constructed from the full
$\Lsl(M+1)$ generators. Therefore, each Hamiltonian of the five cases
classified here must fall within one of the classes of the $\Lsl(M+1)$
Lie-algebraic Hamiltonians classified in Ref.~\cite{Ta04}. To see
the correspondence, it is convenient to extend the free parameters in
the Hamiltonians to be complex-valued and to consider the $GL(2,\bbC)$
of complex linear fractional transformations in Eq.~\eqref{eq:fract}.
\begin{table}
\caption{Correspondence between type A and type C canonical forms}
\label{tb:table2}
\vspace*{.6cm}
\begin{center}
\begin{tabular}{llll}
\hline
\multicolumn{2}{l}{Type A} & $GL(2,\bbC)$ & Type C\\
Case & Canonical Form & \raisebox{-1.2ex}{$\overrightarrow{
 \phantom{AAAAA}}$} & Case\\
\hline
I   & $1/2$ & ---     & --- \\
II  & $2z$  & Trivial & 1\\
III & $2z^2$ & Trivial & 2\\ 
IV  & $2(z^2-1)$ & $\alpha=-2\beta=-2\delta$ & 3($m=0$) \\
    &       & $\phantom{\alpha}=-\sqrt{2}\,\ee^{\ii\pi/4}$, $\gamma=0$ & \\
V   & $2z^3-g_2 z/2-g_3 /2$\quad & $\beta=e_1$, $\gamma=0$, $\delta=1$
 & 3($m\neq 0$), 4, 5\\
\hline 
\end{tabular}
\end{center}
\vspace*{.6cm}
In this Table, $g_2,g_3\in\bbC$ satisfy $g_2^3-27g_3^2\neq 0$, and
$e_1\in\bbC$ is one of the three different roots $e_i$ ($i=1,2,3$)
of $4\beta^3-g_2 \beta-g_3=0$.
\noindent
\end{table}

In Table~\ref{tb:table2}, the first and second columns show the
classification scheme and the canonical form of the polynomials $A(z)$
under $GL(2,\bbC)$, respectively, for the $\Lsl(M+1)$ Lie-algebraic
Hamiltonians preserving one type A space, the third column shows
the concrete set of the parameters of the $GL(2,\bbC)$ transformation
which renders the each canonical form of the type A to one of the canonical
form of the type C, and the fourth column shows the corresponding type C
case classified in this article.
For example, we can read from Table~\ref{tb:table2} that Case 3 with $m=0$
of the type C is equivalent to Case IV of the type A under the $GL(2,\bbC)$
transformation while all of Case 3($m\neq 0$), 4, and 5 are to Case V.

Combining the results here with those in Ref.~\cite{Ta04}, we finally obtain
Table~\ref{tb:table3}. In Table~\ref{tb:table3}, the number 1 or 2 indicates
the number of different families of polynomial type solutions available for
each model
while the symbol $\times$ means that the corresponding model is only
quasi-solvable but not solvable. It is interesting to note that quantum
(quasi-{})solvability favors the integrable models associated with the $BC$
root system over those associated with the $A$ root system. In the
construction of the type C models, on the other hand, we have not assumed
\emph{a priori} the invariance under the Weyl group of the $BC$ root system.
In this respect, we have not appreciated whether this unbalance comes from
a deeper mathematical structure.
\begin{table}
\caption{Classification of the (quasi-{})solvable quantum many-body systems}
\label{tb:table3}
\vspace*{.6cm}
\begin{center}
\begin{tabular}{lcc}
\hline
$\bullet\,$Class ($\cdot\,$Subclass) & Quasi-solvable & Solvable\\
\hline
$\bullet\,$Rational $A$ Inozemtsev     & 1 & $\times$\\
\quad$\cdot\,$Rational $A$ Calogero-Sutherland & 1 & 1\\
$\bullet\,$Rational $BC$ Inozemtsev    & 1 & $\times$\\
\quad$\cdot\,$Rational $BC$ Calogero-Sutherland & 2 & 2\\ 
$\bullet\,$Hyp.(Trig.) $A$ Inozemtsev  & 1 & $\times$\\
\quad$\cdot\,$Hyp.(Trig.) $A$ Calogero-Sutherland & 1 & 1\\
\quad\phantom{$\cdot\,$}+\,external Morse potential &   & \\
$\bullet\,$Hyp.(Trig.) $BC$ Inozemtsev & 1 & $\times$\\
\quad$\cdot\,$Hyp.(Trig.) $BC$ Calogero-Sutherland & 2 & 2\\
$\bullet\,$Elliptic $BC$ Inozemtsev    & 1 & $\times$\\
\quad$\cdot\,$Elliptic $BC$ Calogero-Sutherland & 1 & $\times$\\
\quad$\cdot\,$Jacobi-elliptic $BC$ Inozemtsev & 2 & $\times$\\
\hline 
\end{tabular}
\end{center}
\noindent
\end{table}

As was discussed in the previous paper \cite{GT04a}, one of the two different
families of polynomial type solutions in most cases does not satisfy
the normalizability on a space where the Hamiltonian can be naturally
defined and
thus does not provide a physical state. Even in the case where both of them
can satisfy the normalizability on this space,
one of them is still weakly singular at $q_i=0$ and thus it is natural to
discard it as unphysical. As far as the author's knowledge, however, there
has been no general principle in quantum mechanics which enforces us to do
it\footnote{One of the most acceptable postulates could be that in
Ref.~\cite{LL77}.}. Indeed, some recent investigations have rather indicated
the significance of the careful treatment of this kind of solution, cf.
Refs.~\cite{TFC03,Zn03} and references cited therein. In particular,
it was shown in Ref.~\cite{TFC03} that by employing an ingenious
boundary condition at the singularity of the potential proper linear
combinations of the two linearly independent normalizable solutions, which
still have a weak singularity, may be well-defined as physical eigenstates
of the Hamiltonian.
Therefore, the possibility of realizing such a novel physical state would
deserve further investigations both theoretical and experimental.

Although we have mainly concentrated on the construction of quantum many-body
Hamiltonians in this article, the procedure is quite general and may have
wider applicability. We may construct, for instance, a higher-order linear
differential operator which admits two or more linearly independent series
or polynomial type solutions by a slight modification of the present approach.

Finally, it should be pointed out that the manipulation employed in
Section~\ref{sec:reduc} is justified only after we recognize that the most
general type C quasi-solvable operator preserves the two subspaces in
Eq.~\eqref{eq:mtypC} separately.
In other words, an analogous condition to Eq.~\eqref{eq:tCcon}
in general provides only a sufficient condition for an operator to preserve
the type C space. In fact, we have found that, without the restriction
on the value of the parameter $\lambda$ given by Eq.~\eqref{eq:resla},
the most general type C operator no longer preserves the two subspaces
separately. The restriction \eqref{eq:resla} was originally
introduced~\cite{GT04a} in order to prevent the type C space \eqref{eq:stypC}
of a single variable from reducing to the type A space \eqref{eq:stypA}.
In the several-variable case, however, it is easily seen that the generalized
type C space \eqref{eq:mtypC} does not reduce to the type A space
\eqref{eq:mtypA} of several variables even if the parameter $\lambda$
takes a forbidden value in Eq.~\eqref{eq:resla}. Therefore, there exists
the cases, which we shall refer to as \emph{irregular cases}, where
the monomial space is of the type C and different from the type A but
nevertheless the general quasi-solvable operator for it does not preserve
the two subspaces separately. These irregular cases need separate treatment
from the regular ones presented in this article and will be reported
elsewhere.


\begin{ack}
  The author would like to thank A.~Gonz\'alez-L\'opez for the useful
  discussion.
  This work was partially supported by a Spanish Ministry of
  Education, Culture and Sports research fellowship.
\end{ack}

\appendix

\section{Formulas}

In this appendix, we summarize the formulas connecting the differential
operators in terms of $z_i$ with those in terms of $\sigma_k$. For the
details of the derivation, see Ref.~\cite{Ta04}.
\begin{subequations}
\label{eqs:form1}
\begin{align}
\sum_{i=1}^M z_i \frac{\del^2}{\del z_i^2}=&\,-\sum_{k,l=1}^M
 \biggl[\,\sum_{m=0}^{k-1}(k-l-2m-1)\sigma_{k-m-1}\sigma_{l+m}\biggr]
 \frac{\partial^2}{\partial\sigma_k \partial\sigma_l}\,,\\
\sum_{i=1}^M z_i^2 \frac{\del^2}{\del z_i^2}=&\,\sum_{k,l=1}^M
 \biggl[ k\,\sigma_k \sigma_l +\sum_{m=1}^k (k-l-2m)
 \sigma_{k-m}\sigma_{l+m}\biggr] \frac{\partial^2}{\partial\sigma_k
 \partial\sigma_l}\,,\\
\sum_{i=1}^M z_i^3 \frac{\del^2}{\del z_i^2}=&\,\sum_{k,l=1}^M
 \biggl[ \sigma_1 \sigma_k \sigma_l -\sum_{m=0}^1 (k-m+1)
 \sigma_{k-m+1}\sigma_{l+m}\notag\\
&\,\qquad -\sum_{m=2}^{k+1}(k-l-2m+1)\sigma_{k-m+1}\sigma_{l+m}
 \biggr]\frac{\partial^2}{\partial\sigma_k \partial\sigma_l}\,.
\end{align}
\end{subequations}
\begin{subequations}
\label{eqs:form2}
\begin{align}
\sum_{i=1}^M \frac{\del}{\del z_i}=&\,
 -\sum_{k=1}^M (k-M-1)\sigma_{k-1}\frac{\del}{\del\sigma_k}\,,\\
\sum_{i=1}^M z_i \frac{\del}{\del z_i}=&\,
 \sum_{k=1}^M k\,\sigma_k \frac{\del}{\del\sigma_k}\,,\\
\sum_{i=1}^M z_i^2 \frac{\del}{\del z_i}=&\,
 \sum_{k=1}^M \bigl[\sigma_1 \sigma_k -(k+1)\sigma_{k+1}\bigr]
 \frac{\del}{\del\sigma_k}\,.
\end{align}
\end{subequations}
\begin{subequations}
\label{eqs:form3}
\begin{align}
2\sum_{i\neq j}^M \frac{z_i}{z_i-z_j}\frac{\del}{\del z_i}=&\,
 \sum_{k=1}^M (k-M)(k-M-1)\sigma_{k-1}
 \frac{\del}{\del\sigma_k}\,,\\
2\sum_{i\neq j}^M \frac{z_i^2}{z_i-z_j}\frac{\del}{\del z_i}=&\,
 -\sum_{k=1}^M k(k-2M+1)\sigma_k \frac{\del}{\del\sigma_k}\,,\\
2\sum_{i\neq j}^M \frac{z_i^3}{z_i-z_j}\frac{\del}{\del z_i}=&\,
 \sum_{k=1}^M \bigl[ 2(M-1)\sigma_1 \sigma_k \notag\\
&\,\qquad +(k+1)(k-2M+2)\sigma_{k+1}\bigr]
 \frac{\del}{\del\sigma_k}\,.
\end{align}
\end{subequations}

\bibliography{refs}
\bibliographystyle{npb}



\end{document}